\documentclass[10pt]{article}
\usepackage{a4}
\usepackage{graphicx}
\newcommand{\vect}[1]{\mbox{${\bf #1}$}}
\newcommand{\gvect}[1]{\mbox{\boldmath$ #1$}}

\newcommand{\vdot}{\mbox{\boldmath $\cdot$}}

\begin{document}

\title{Nuclear applications of inverse scattering,\\ present \ldots\
and future? }
\author{R.S. Mackintosh \\Department of Physics and Astronomy\\
 The Open University\\ MK7 6AA, UK}
\date{  }
\maketitle

\begin{abstract}
There now exists a practical method (IP) for the routine inversion
of $S$-matrix elements to produce the corresponding
potential~\cite{km}. It can be applied to spin-$\frac{1}{2}$
 spin-1 projectiles. We survey the ways that the
method can be applied in nuclear physics, by inverting $S_{lj}$
derived from theory or from experiment. The IP method can be
extended to invert $S_{lj}(E)$ over a range of energies to produce
a potential $V(r,E) + \vect{l}\vdot\gvect{\sigma} V_{\rm
ls}(r,E)$, yields parity-dependent potentials between pairs of
light nuclei~\cite{rm742} and can be convoluted with a direct
search on the $S$-matrix to produce `direct data $\rightarrow V$
inversion'. The last is an economical alternative form of optical
model search to fit many observables (e.g. for polarized
deuterons) for many energies, producing an energy-dependent
potential with many parameters (e.g. $T_{\rm R}$ for
deuterons)~\cite{cm723}.

\end{abstract}

\section{Introduction}
It is very easy to derive a cross section or an $S$-matrix from a
potential; the reverse is much harder, but can now routinely be
achieved for a wide range of cases. Here we introduce the IP
(iterative-perturbative) inverse scattering method, with the
emphasis on its range of applicability, illustrated by a few
diverse successful applications. We hope to inspire applications
that we have not thought of.

Alternative methods exist for some tasks to which IP inversion can
be applied (e.g.\ conventional OM searches to obtain potentials
from scattering data, and weighted trivially equivalent potentials
(TELPs) to derive DPPs) but IP inversion not only has advantages
in these cases but also gives reliable results where no other
techniques are available.

\section{Inverse scattering: problems and solutions}
`Inverse scattering' usually refers to the derivation of the
potential corresponding to given $S$-matrix elements or phase
shifts; it may be contrasted with the trivial forward case. This
is Case 1 of Table~\ref{trad}. We shall say little about Case 2 of
the table, but we will mention Case 3, direct inversion from
observables to potential, since the IP method can readily be
convoluted with Case 2 inversion to yield a very powerful Case 3
alternative to conventional optical model searches.
\begin{table}[htb] \begin{center}
\quad \begin{tabular}{|l|l|c|c|} \hline
Case 1& forward&$V(r) \rightarrow S_l$ & EASY\\
& inverse&$S_l \rightarrow V(r)$\ & MUCH HARDER\\ \hline
Case 2&forward&$S_l \rightarrow \sigma(\theta)$ & TRIVIAL\\
& inverse&$\sigma(\theta) \rightarrow S_l$ & OFTEN VERY HARD \\ \hline
Case 3&forward&$V(r) \rightarrow \sigma(\theta)$ & EASY\\
& inverse&$\sigma(\theta) \rightarrow V(r)$ & VERY HARD\\ \hline \end{tabular}
\caption{Traditional characterization of forward and inverse scattering cases. The $\sigma$ symbolizes
all observables (analyzing powers, etc); $S_l$ includes $S_{lj}$ etc;  $V$ includes
spin-orbit and tensor terms.}\label{trad}
 \end{center} \end{table}
\subsection{Traditional methods of $S_l \rightarrow V(r)$ inversion}
Formal inversion methods apply to two classes:\\
1. {\bf Fixed-$l$ inversion.} (Gel'fand and Levitan; Marchenko)
$S_l\rightarrow V(r)$ for $S_l$ for a single $l$ for {\em all\/} energies,
yielding a local potential.\\
2. {\bf Fixed energy inversion.} (Newton-Sabatier (NS); M\"unchow
and Scheid (MS)) $S_l \rightarrow V(r)$ given $S_l$ for {\em
all\/} $l$ at a single energy. Related procedures have been developed by
Lipperheide and Fiedeldey.\\
There also exist semi-classical methods for fixed-energy inversion
based on WKB and Glauber (eikonal) methods. Practical versions of
NS, MS, can handle a finite range of $l$. The review~\cite{km}
gives comprehensive references to the inversion techniques.

{\bf  Disadvantages of traditional methods:}\\
(1) The NS method requires highly precise $S_l$, and there is a tendency to instability.\\
(2)  The Marchenko method requires large energy ranges, but nuclear potentials
are energy-dependent (but it has been used for nucleon-nucleon scattering). \\
(3)  They are not adaptable to cases with small ranges of $l$ (NS).\\
(3)  They mostly apply to spin 0; NS can handle spin 1/2. \\
(4) They are not readily generalizable.\\
Nevertheless, there have been a number of applications which have yielded
real physical insights, see Ref.~\cite{km}.
\subsection{Information on nuclear interactions from inversion}
Inversion can be applied in three distinct general ways:\\[2 mm]
{\bf I. Inversion of $S_l$, $S_{lj}$ or $S_{ll'}^j$ obtained from theory:}
\begin{enumerate}
\item Derive dynamic polarization potentials (DPPs) arising from
(i) Inelastic scattering, (ii) Breakup processes, (iii)
Reaction channels, etc.
\item Derive potential  from RGM and similar $S$-matrix elements.
\item Determine local potentials $S$-matrix equivalent to
 non-local potentials.
\item Obtain a potential representation of impulse approximation $S$-matrix,
or the $S$-matrix from Glauber model and other impact parameter models.
\end{enumerate}
{\bf II. Inversion of $S_l$, $S_{lj}$ or $S_{ll'}^j$ from analysis of
 experiment:}
\begin{enumerate} \item (When there are few partial waves.) Invert $S_l$ from
parameterized $R$-matrix or effective range fits at low energies. (Requires
`mixed case' or `energy dependent' inversion, see Section 3.3)
\item (When there are many partial
waves) High energy two-step phenomenology. (E.g. $^{11}$Li, $^{12}$C +$^{12}$C from
140 to 2400 MeV)
\end{enumerate}
{\bf III. Direct observable $\rightarrow V(r)$ inversion.} $S$-matrix
search can be convoluted with IP inversion yielding the $S$-matrix is byproduct. Many
energies can be treated simultaneously to give a multi-component $V(E)$.

 \section{The Iterative-Perturbative (IP) method}
\subsection{The key idea} The response of the elastic scattering $S$-matrix to
small changes is assumed to be linear (this is often surprisingly accurate):
\begin{equation}  \Delta S_l = -\frac{{\rm i} m}{\hbar^2 k}
\int_0^{\infty} (u_l(r))^2 \Delta V(r) {\rm d}r. \label{delta} \end{equation} where
$u_l(r)$ is normalized with $u_l(r) \rightarrow I_l(r) - S_l O_l(r)$,
$I_l$ and $O_l$ are incoming and outgoing Coulomb wavefunctions. Eq.~\ref{delta}
is readily generalized for spin.

\subsection{An Outline of the IP method:}
Take a known `starting reference potential', SRP, $V(r)$ giving
$S_l$. With added term:
 \begin{equation} V(r) \rightarrow \hat{V}(r)= V(r) + \sum c_i v_i(r) \label{basic}
\end{equation} it gives $S_l+ \Delta
S_l$. Functions $v_i(r)$ belong to a suitable `inversion basis'.

The core of the method is the solution using SVD  of the
over-determined linear equations derived from Eqn.~\ref{delta}
with $\Delta S_l =S_l^{\rm target} -S_l$ and $\Delta V =\sum c_i
v_i(r)$ to find amplitudes $c_i$ such that $\hat{V}$ gives $S_l$
closer to $S_l^{\rm target}$. By iterating the linear equations,
$S_l+ \Delta S_l$ converges to $S_l^{\rm target}$. There is often
a natural starting potential; it can often be zero, or the `bare
potential' when establishing DPP contributions.

The facility of the IP method to control the fitting is a key
element in the method, especially in view of the innate
ill-posedness (Ref.~\cite{leeb}) of the inversion problem; it is
possible always to demand smooth potentials. This and all aspects
are fully discussed in Ref.~\cite{km} with many references.

\subsection{Generalizing from fixed-energy inversion}
IP Inversion can be generalized indefinitely as the following progression suggests:\\
{\bf 1. Fixed-energy inversion}  \underline{$S_l$, `all $l$, one $E$'} inversion. However,
at low energies the  potential is generally under-determined, there being too few
active partial
waves to define the required potential.\\
{\bf 2. Mixed case (energy bite) inversion.}
The problem of under-determination at low energies can often be solved given
$S_l(E)$ over a range of energies (`energy bite'). This is
\underline{`some $l$, some $E$', $S_l(E)\rightarrow V(r)$}, or `mixed case', inversion.
 For a narrow energy bite, this is effectively
includes ${\rm d}S_l/{\rm d}E$ as input information.\\
{\bf 3. Energy dependent inversion.} Nuclear potentials,
 particularly the imaginary parts, vary with energy, but  IP
inversion can be extended to determine $V(r,E)$ directly:
\underline{`some $l$, some $E$', $S_l(E)\rightarrow V(r,E)$}.
To date, cases have been limited to the factored form $\sum_i f_i(E)V_i(r)$,
where $i$ indicates real-central, imaginary-central, spin-orbit, Majorana, etc.\\
{\bf 4. Inversion to fit bound state and resonance energies.} Energies
of bound states can be included with $S_l$ as input information to determine $V$.\\
{\bf 5. Direct data to potential inversion.} Example: $\vec{\rm d} - ^4$He,
multi-energy.

IP inversion can be applied to the case of identical bosons where
only even partial waves are involved, e.g.\ $^{12}$C + $^{12}$C.
It also applies to the case, very important with light nuclei,
where the potential is parity dependent. Such a potential can
either be represented as a sum of independent even-parity and
odd-parity terms, or as a sum of Wigner (W) and Majorana (M)
terms: \[ V_{\rm W}(r) + (-1)^L V_{\rm M}(r). \] It is always
found that even-parity and odd-parity potentials have different
radial forms which often implies a surface peaked Majorana term.
Both RGM $S$-matrices and experimental data imply that there is
significant parity dependence even for nucleons on nuclei as heavy
as $^{16}$O.

\subsection{Spin cases that can be handled by IP inversion}
{\bf 1. Spinless projectiles.}
$S_l \rightarrow V(r)$.\\
{\bf 2. Spin 1/2 projectiles.} $S_{lj} \rightarrow V(r) + {\bf l} \cdot
 \sigma V_{ls}(r)$.\\
{\bf 3. Spin one projectiles.} Vector  spin-orbit and
  $T_{\rm R}\equiv ({\bf (s\cdot \hat{r})^2} -2/3) V_R(r)$ tensor
potentials can be determined from non-diagonal $S^j_{ll'}$. This is
{\em coupled channel inversion\/}.\\
{\bf 4. High channel spin.} For cases like d + $^3$He, independent potentials
for each possible channel spin have been determined.

In every case, all spin-dependent components may have real and
imaginary and Wigner and Majorana terms. These can all be expanded
in different bases.

\subsection{How well does it work?}\label{how-well}
Fig.~1 shows a test case~\cite{npa677} in which a $S_{ll'}^j$ for
deuterons on $^{58}$Ni at 56 MeV and a known potential, including
a tensor term, were inverted with an arbitrary SRP.  IP inversion
can be applied to noisy data and produce meaningful interactions
because the departure of the final potential from the `starting
potential' of the iterative method is under control, see Section
4.3.
\begin{figure}[ht]
\begin{center}
\includegraphics[scale=.5]{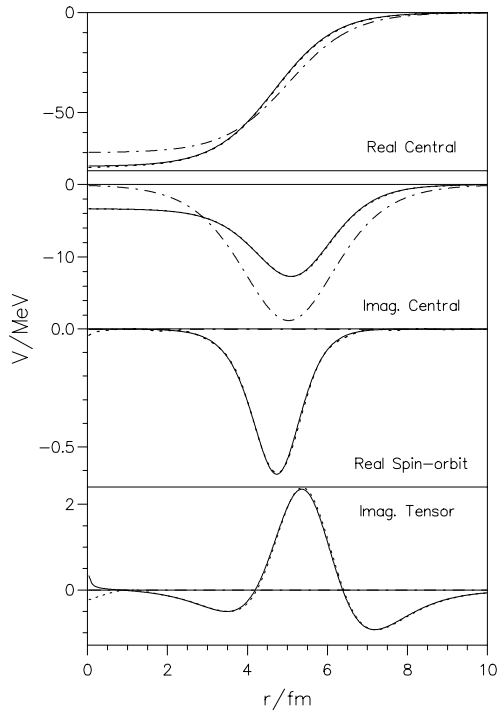}
\end{center}
\label{test} \caption{Deuterons on $^{58}$Ni at 56 MeV. Solid
lines: target (known) potential; dots: $V$ found by inversion;
dash-dot: inversion SRP (starting potentials, zero for the real
spin-orbit and tensor terms.)}
\end{figure}

\section{Selected applications of IP inversion}
{\bf Light nuclei, parity dependence.}\quad Scattering between
various pairs of light nuclei have been studied by inverting
$S$-matrices from  RGM calculations and from $R$-matrix fits to
experimental data, over a wide range of energies. These studies
reveal the importance of a parity-dependent (Majorana) component.
Ref.~\cite{km} has references to parity-dependent potentials for
various pairs of light nuclei and presents a case study of $p +
^4$He, comparing potentials from empirical and theoretical
$S_{lj}$. The contrasting Majorana potential for $d + ^4$He is
discussed in Ref.~\cite{rm728} and that for $p + ^6$He in
Ref.~\cite{rm742}.

{\bf The dynamic polarization potential, DPP.} It is well known
that the coupling to breakup channels generates a repulsive DPP
for projectiles such as $^6$Li and $^2$H. Inverting the elastic
scattering $S$-matrix from a coupled channel calculation, and
subtracting the bare potential, gives a local-equivalent
$l$-independent representation of the DPP. The form of the DPP
depends on the $L$-transfer in a systematic way~\cite{im}.
Nucleus-nucleus interaction also receive large contributions from
coupling to inelastic and (especially) reaction channels that {\em
cannot be represented by renormalizing folding model potential\/.}
Many cases are described in Ref.~\cite{km}, and the contribution
of breakup to the $p + ^6$He interaction is presented in
Ref.~\cite{rusek}, revealing the limitations of folding models for
halo nuclei.

{\bf Potentials from empirical data.}\quad Elastic scattering
potentials, including $^{12}$C + $^{12}$C, $^{16}$O + $^{16}$O,
$^{11}$Li scattering and nucleon scattering, have been determined
either by fitting $S$ to data and then inverting, or by using the
'direct inversion' in which the $S$-matrix search is convoluted
with the $S\rightarrow V$ inversion. The definitive phenomenology
for $p + ^{16}$O has been carried out in this
way~\cite{cooper618}, revealing, {\em inter alia\/}, the necessity
for a Majorana term. Direct inversion of multi-energy data for
scattering of light nuclei provides an alternative means of
establishing phase shifts $\delta_{lj}(E)$ that behave in a way
that is consistent with a potential~\cite{skmk} that varies
smoothly with energy. Inversion of $S_{lj}$ from R-matrix fits
also has this property.

\subsection{Pickup coupling effect in $^8$He($p,p$) elastic
scattering} It known that pickup coupling ($p\rightarrow d
\rightarrow p$ for proton scattering) makes a significant
contribution to the nucleon optical potential. This is one reason
that precise fits for nucleon elastic scattering below 50 MeV for
closed shell target nuclei have not been found with conventional
optical model fitting (for non-closed shell nuclei it is easier to
find parameters that fit the shallower diffraction minima). It is
now possible to do full finite-range pickup calculations including
non-orthogonality terms, and these have been carried
out~\cite{skaza} for the p-$^8$He system. The coupling has a large
effect on the elastic scattering angular distributions. In Table~2
we quantify the pickup contribution in terms of volume integrals.
The repulsive real DPP is quite large at the nuclear centre
although the effect on the volume integral is modest. The radial
form of the DPP could {\em not\/} be represented by renormalizing
a folding model potential. The volume integrals reveal the
importance of including the non-orthogonality correction.
\begin{table}[h]
\begin{center}
\begin{tabular}{||l|r|r|r|r|r|r|r||}\hline
&$J_{\rm R}$& $\langle r^2\rangle_{\rm R}^{1/2}$& $J_{\rm I}$& $\langle
r^2\rangle_{\rm I}^{1/2}$&$J_{\rm SOR}$& $J_{\rm SOI}$\\ \hline
OM &  704.14 & 3.092 & 55.37 & 3.336 & 26.60 & 0.005\\  \hline
CRC & 653.94 & 2.938 & 307.47 &4.138 &40.27& 1.25 \\   \hline
NONO & 571.28 &2.840 &252.62 &4.360 &33.15 & 6.55 \\ \hline
\end{tabular}
\end{center}\label{phe8}
\caption{For 15.6 MeV protons on $^8$He, volume integrals per
nucleon pair/(MeV fm$^3$), and rms radii/fm of the bare potential
(OM) and the potentials found by inversion for the complete CRC
calculation and for that in which non-orthogonality term was
omitted (NONO).}
\end{table}

\subsection{The DPP due to breakup for $^6$He scattering from $^{208}$Pb.}
The breakup for this case~\cite{rmnk,nkrm} was calculated using
the same model for $^6$He as Ref.~\cite{rusek}. However the DPP is
now very different, having a long range attractive tail generated
by the Coulomb dipole interaction. Fig.~2 shows the short range
repulsive/emissive DPP in the surface region. Local regions of
emissiveness are a common feature of local potentials representing
the highly non-local and $l$-dependent dynamical polarization
contributions; unitarity is not broken. The DPP is not
well-defined for $r\le 10.5$ fm.
\begin{figure}[ht]
\begin{center}
\includegraphics[scale=.45]{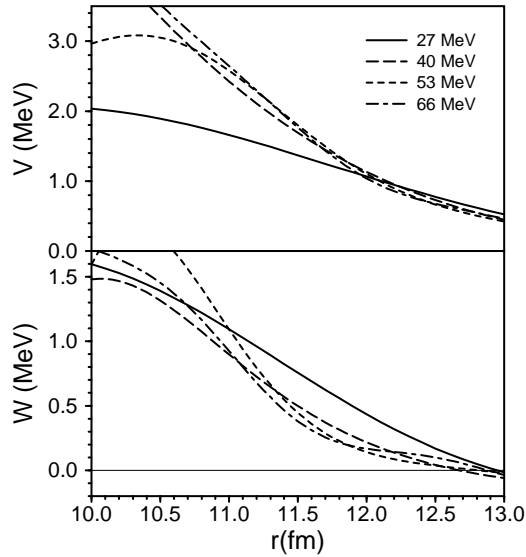}
\end{center}
\label{pbhe6-1} \caption{For
 $^6$He incident on $^{208}$Pb, 27 to 66 MeV, the  DPP at the nuclear surface.}
\end{figure}

Fig.~3 shows the long range attractive and absorptive potentials
generated by the dipole coupling; the real part is appreciable out
to 60 fm. Both the real and imaginary DPPs strongly influence the
elastic scattering differential cross section.

\begin{figure}[ht]
\begin{center}
\includegraphics[scale=.45]{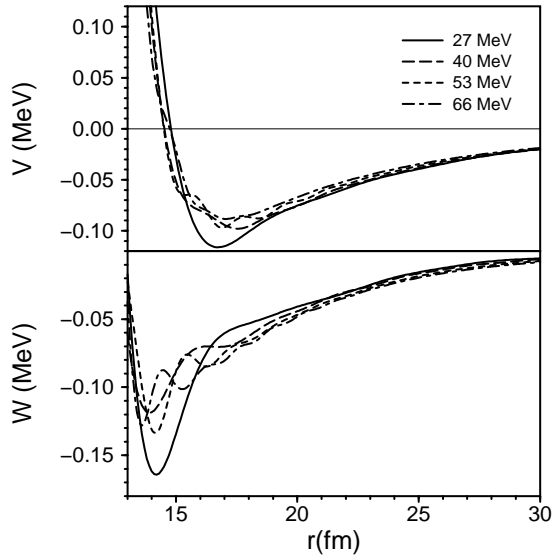}
\end{center}
\label{pbhe6-2} \caption{For
 $^6$He incident on $^{208}$Pb, 27 to 66 MeV, the  DPP outside the nucleus.}
\end{figure}

\subsection{The $d$-$^4$He interaction derived from multi-energy
data} `Direct data $\rightarrow V$ inversion' is an
alternative~\cite{skmk} to optical model fitting with
parameterized forms for determining potentials from data,
especially when there are many data and many parameters. This is
the case for $d$-$^4$He scattering when there is a full set of
polarization observables (including all 3 tensor analyzing powers)
for many angles, all for many energies ranging from 4 to 13 MeV.
Refs.~\cite{cm723} fitted 1000 data points (five observables, a
wide angular range and many energies) to produce a multi-component
(Wigner and Majorana, central, spin-orbit and tensor) multi-energy
potential (components were functions of energy) giving a
reasonable representation of shape resonances. This would have
been a formidable task for standard optical model codes since they
would have had to include the coupled channel calculation for the
tensor interaction within the search.

\section{Possible future applications}
1. Systematic CRC calculations followed by inversion of the
resulting elastic $S_{lj}$ provides a
method for establishing shell corrections to the nucleon OM potential.\\
2. There exists much data for the elastic scattering of neutrons
from $^{12}$C. Inversion of phase shifts fitted to this data would
yield a potential model fitting the non-resonant scattering,
allowing an extrapolation to higher neutron energies. Comparing
the potential derived in this way with theoretical and standard
empirical OM potentials would support (or the opposite) the
neutron scattering data as well as the theoretical models.\\
3. The IP algorithm appears to be indefinitely generalizable, and
has found wider applications than originally envisaged, so are
there new applications to nuclear data evaluation? Any
suggestions?


\end{document}